\newcommand\blfootnote[1]{%
\begingroup
\renewcommand\thefootnote{}\footnote{#1}%
\addtocounter{footnote}{-1}%
\endgroup
}

\documentclass[prd, superscriptaddress, twocolumn, reprint, preprintnumbers, nofootinbib]{revtex4-2}

\input{header}

\begin{document}



\preprint{UCI-TR-2021-04, KYUSHU-RCAPP-2020-04, CERN-EP-2021-087}

\title{First neutrino interaction candidates at the LHC}

\author{Henso Abreu}
\affiliation{Department of Physics and Astronomy, Technion---Israel Institute of Technology, Haifa 32000, Israel}

\author{Yoav Afik}
\affiliation{Department of Physics and Astronomy, Technion---Israel Institute of Technology, Haifa 32000, Israel}

\author{Claire Antel}
\affiliation{D\'epartement de Physique Nucl\'eaire et Corpusculaire, 
University of Geneva, CH-1211 Geneva 4, Switzerland}

\author{Jason Arakawa}
\affiliation{Department of Physics and Astronomy, 
University of California, Irvine, CA 92697-4575, USA}

\author{Akitaka Ariga}
\affiliation{Albert Einstein Center for Fundamental Physics, Laboratory for High Energy Physics, University of Bern, Sidlerstrasse 5, CH-3012 Bern, Switzerland}
\affiliation{Department of Physics, Chiba University, 1-33 Yayoi-cho Inage-ku, Chiba, 263-8522, Japan}

\author{Tomoko Ariga}
\email[Corresponding author: ]{tomoko.ariga@cern.ch}
\affiliation{Kyushu University, Nishi-ku, 819-0395 Fukuoka, Japan}

\author{Florian Bernlochner}
\affiliation{Universit\"at Bonn, Regina-Pacis-Weg 3, D-53113 Bonn, Germany}

\author{Tobias Boeckh}
\affiliation{Universit\"at Bonn, Regina-Pacis-Weg 3, D-53113 Bonn, Germany}

\author{Jamie Boyd}
\affiliation{CERN, CH-1211 Geneva 23, Switzerland}

\author{Lydia Brenner}
\affiliation{CERN, CH-1211 Geneva 23, Switzerland}

\author{Franck Cadoux}
\affiliation{D\'epartement de Physique Nucl\'eaire et Corpusculaire, 
University of Geneva, CH-1211 Geneva 4, Switzerland}

\author{David~W.~Casper}
\affiliation{Department of Physics and Astronomy, 
University of California, Irvine, CA 92697-4575, USA}

\author{Charlotte Cavanagh}
\affiliation{University of Liverpool, Liverpool L69 3BX, United Kingdom}

\author{Francesco Cerutti}
\affiliation{CERN, CH-1211 Geneva 23, Switzerland}

\author{Xin Chen}
\affiliation{Department of Physics, Tsinghua University, Beijing, China}

\author{Andrea Coccaro} 
\affiliation{INFN Sezione di Genova, Via Dodecaneso, 33--16146, Genova, Italy}

\author{Monica D’Onofrio} 
\affiliation{University of Liverpool, Liverpool L69 3BX, United Kingdom}

\author{Candan Dozen}
\affiliation{Department of Physics, Tsinghua University, Beijing, China}

\author{Yannick Favre}
\affiliation{D\'epartement de Physique Nucl\'eaire et Corpusculaire, 
University of Geneva, CH-1211 Geneva 4, Switzerland}

\author{Deion Fellers}
\affiliation{University of Oregon, Eugene, OR 97403, USA}

\author{Jonathan~L.~Feng}
\affiliation{Department of Physics and Astronomy, 
University of California, Irvine, CA 92697-4575, USA}

\author{Didier Ferrere}
\affiliation{D\'epartement de Physique Nucl\'eaire et Corpusculaire, 
University of Geneva, CH-1211 Geneva 4, Switzerland}

\author{Stephen Gibson}
\affiliation{Royal Holloway, University of London, Egham, TW20 0EX, UK}

\author{Sergio Gonzalez-Sevilla}
\affiliation{D\'epartement de Physique Nucl\'eaire et Corpusculaire, 
University of Geneva, CH-1211 Geneva 4, Switzerland}

\author{Carl Gwilliam}
\affiliation{University of Liverpool, Liverpool L69 3BX, United Kingdom}

\author{Shih-Chieh Hsu}
\affiliation{Department of Physics, University of Washington, PO Box 351560, Seattle, WA 98195-1560, USA}

\author{Zhen Hu}
\affiliation{Department of Physics, Tsinghua University, Beijing, China}

\author{Giuseppe Iacobucci}
\affiliation{D\'epartement de Physique Nucl\'eaire et Corpusculaire, 
University of Geneva, CH-1211 Geneva 4, Switzerland}

\author{Tomohiro Inada}
\affiliation{Department of Physics, Tsinghua University, Beijing, China}

\author{Ahmed Ismail}
\affiliation{Oklahoma State University, Stillwater, OK 74078-3072, USA}

\author{Sune Jakobsen}
\affiliation{CERN, CH-1211 Geneva 23, Switzerland}

\author{Enrique Kajomovitz}
\affiliation{Department of Physics and Astronomy, 
Technion---Israel Institute of Technology, Haifa 32000, Israel}

\author{Felix Kling}
\affiliation{SLAC National Accelerator Laboratory, 2575 Sand Hill Road, Menlo Park, CA 94025, USA}

\author{Umut Kose}
\affiliation{CERN, CH-1211 Geneva 23, Switzerland}

\author{Susanne Kuehn}
\affiliation{CERN, CH-1211 Geneva 23, Switzerland}

\author{Helena Lefebvre}
\affiliation{Royal Holloway, University of London, Egham, TW20 0EX, UK}

\author{Lorne Levinson}
\affiliation{Department of Particle Physics and Astrophysics, Weizmann Institute of Science, Rehovot 76100, Israel}

\author{Ke Li}
\affiliation{Department of Physics, University of Washington, PO Box 351560, Seattle, WA 98195-1560, USA}

\author{Jinfeng Liu}
\affiliation{Department of Physics, Tsinghua University, Beijing, China}

\author{Chiara Magliocca}
\affiliation{D\'epartement de Physique Nucl\'eaire et Corpusculaire, 
University of Geneva, CH-1211 Geneva 4, Switzerland}

\author{Josh McFayden}
\affiliation{Department of Physics \& Astronomy, University of Sussex, Sussex House, Falmer, Brighton, BN1 9RH, United Kingdom}

\author{Sam Meehan}
\affiliation{CERN, CH-1211 Geneva 23, Switzerland}

\author{Dimitar Mladenov}
\affiliation{CERN, CH-1211 Geneva 23, Switzerland}

\author{Mitsuhiro Nakamura}
\affiliation{Nagoya University, Furo-cho, Chikusa-ku, Nagoya 464-8602, Japan}

\author{Toshiyuki Nakano}
\affiliation{Nagoya University, Furo-cho, Chikusa-ku, Nagoya 464-8602, Japan}

\author{Marzio Nessi}
\affiliation{CERN, CH-1211 Geneva 23, Switzerland}

\author{Friedemann Neuhaus}
\affiliation{Institut f\"ur Physik, Universität Mainz, Mainz, Germany}

\author{Laurie Nevay}
\affiliation{Royal Holloway, University of London, Egham, TW20 0EX, UK}

\author{Hidetoshi Otono}
\affiliation{Kyushu University, Nishi-ku, 819-0395 Fukuoka, Japan}

\author{Carlo Pandini}
\affiliation{D\'epartement de Physique Nucl\'eaire et Corpusculaire, 
University of Geneva, CH-1211 Geneva 4, Switzerland}

\author{Hao Pang}
\affiliation{Department of Physics, Tsinghua University, Beijing, China}

\author{Lorenzo Paolozzi}
\affiliation{D\'epartement de Physique Nucl\'eaire et Corpusculaire, 
University of Geneva, CH-1211 Geneva 4, Switzerland}

\author{Brian Petersen}
\affiliation{CERN, CH-1211 Geneva 23, Switzerland}

\author{Francesco Pietropaolo}
\affiliation{CERN, CH-1211 Geneva 23, Switzerland}

\author{Markus Prim}
\affiliation{Universit\"at Bonn, Regina-Pacis-Weg 3, D-53113 Bonn, Germany}

\author{Michaela Queitsch-Maitland}
\affiliation{CERN, CH-1211 Geneva 23, Switzerland}

\author{Filippo Resnati}
\affiliation{CERN, CH-1211 Geneva 23, Switzerland}

\author{Hiroki Rokujo}
\affiliation{Nagoya University, Furo-cho, Chikusa-ku, Nagoya 464-8602, Japan}

\author{Marta Sabat\'e-Gilarte}
\affiliation{CERN, CH-1211 Geneva 23, Switzerland}

\author{Jakob Salfeld-Nebgen}
\affiliation{CERN, CH-1211 Geneva 23, Switzerland}

\author{Osamu Sato}
\affiliation{Nagoya University, Furo-cho, Chikusa-ku, Nagoya 464-8602, Japan}

\author{Paola Scampoli}
\affiliation{Albert Einstein Center for Fundamental Physics, Laboratory for High Energy Physics, University of Bern, Sidlerstrasse 5, CH-3012 Bern, Switzerland}
\affiliation{Dipartimento di Fisica ``Ettore Pancini'', Universit\`a di Napoli Federico II, Complesso Universitario di Monte S. Angelo, I-80126 Napoli, Italy}

\author{Kristof Schmieden}
\affiliation{Institut f\"ur Physik, Universität Mainz, Mainz, Germany}

\author{Matthias Schott}
\affiliation{Institut f\"ur Physik, Universität Mainz, Mainz, Germany}

\author{Anna Sfyrla}
\affiliation{D\'epartement de Physique Nucl\'eaire et Corpusculaire, 
University of Geneva, CH-1211 Geneva 4, Switzerland}

\author{Savannah Shively}
\affiliation{Department of Physics and Astronomy, 
University of California, Irvine, CA 92697-4575, USA}

\author{John Spencer}
\affiliation{Department of Physics, University of Washington, PO Box 351560, Seattle, WA 98195-1560, USA}

\author{Yosuke Takubo}
\affiliation{Institute of Particle and Nuclear Study, 
KEK, Oho 1-1, Tsukuba, Ibaraki 305-0801, Japan}

\author{Ondrej Theiner}
\affiliation{D\'epartement de Physique Nucl\'eaire et Corpusculaire, 
University of Geneva, CH-1211 Geneva 4, Switzerland}

\author{Eric Torrence}
\affiliation{University of Oregon, Eugene, OR 97403, USA}

\author{Sebastian Trojanowski}
\affiliation{Astrocent, Nicolaus Copernicus Astronomical Center Polish Academy of Sciences, ul.~Bartycka 18, 00-716 Warsaw, Poland}

\author{Serhan Tufanli}
\affiliation{CERN, CH-1211 Geneva 23, Switzerland}

\author{Benedikt Vormwald}
\affiliation{CERN, CH-1211 Geneva 23, Switzerland}

\author{Di Wang}
\affiliation{Department of Physics, Tsinghua University, Beijing, China}

\author{Gang Zhang}
\affiliation{Department of Physics, Tsinghua University, Beijing, China}

\collaboration{FASER Collaboration}
\noaffiliation

\date{\today}


\begin{abstract}



FASER$\nu$ at the CERN Large Hadron Collider (LHC) is designed to directly detect collider neutrinos for the first time and study their cross sections at TeV energies, where no such measurements currently exist. In 2018, a pilot detector employing emulsion films was installed in the far-forward region of ATLAS, 480 m from the interaction point, and collected 12.2 fb$^{-1}$ of proton-proton collision data at a center-of-mass energy of 13 TeV. We describe the analysis of this pilot run data and the observation of the first neutrino interaction candidates at the LHC. This milestone paves the way for high-energy neutrino measurements at current and future colliders.


\end{abstract}


\maketitle

\blfootnote{\vspace{1mm}}
\blfootnote{\copyright 2021 CERN for the benefit of the FASER Collaboration.}
\blfootnote{Reproduction of this article or parts of it is allowed as specified in the CC-BY-4.0 license.}

\section{Introduction}

There has been a longstanding interest in detecting neutrinos produced at colliders~\cite{DeRujula:1984pg, Winter:1990ry, Vannucci:1993ud, DeRujula:1992sn, Park:2011gh, Feng:2017uoz}, but to date no collider neutrino has ever been directly detected. Proton-proton ($pp$) collisions at a center-of-mass energy of 14 TeV during LHC Run-3, with an expected integrated luminosity of 150 fb$^{-1}$, will produce a high-intensity beam of $\mathcal{O}(10^{12})$ neutrinos in the far-forward direction with mean interaction energy of about 1 TeV.
FASER$\nu$~\cite{Abreu:2019yak} is designed to detect these neutrinos and study their properties. The detector was approved in December 2019, will be installed 480~m downstream of the ATLAS interaction point (IP) in 2021, and will take data starting in 2022. Deployment on the beam collision axis maximizes the flux of all three neutrino flavors, and allows FASER$\nu$ to measure their interaction cross sections in the currently unexplored TeV energy range. For electron and tau neutrinos, these measurements will extend existing cross section measurements to significantly higher energies. For muon neutrinos, they will probe the gap between accelerator measurements ($E_\nu<360$ GeV)~\cite{Zyla:2020zbs} and IceCube data ($E_\nu>6.3$ TeV)~\cite{Aartsen:2017kpd}. Following the FASER$\nu$ approval, the SND@LHC experiment~\cite{Ahdida:2750060}, designed to measure neutrinos at the LHC in a complementary rapidity region to FASER$\nu$, was approved in 2021.

In 2018, we performed a pilot run in the LHC tunnel to measure background and demonstrate neutrino detection at the LHC for the first time. Although the pilot detector lacked the ability to identify muons, given its depth of only 0.6$\lambda_{\text{int}}$, much shorter than the 8$\lambda_{\text{int}}$ of the full FASER$\nu$ detector, these data have aided reconstruction tool development and proven the feasibility of neutrino measurements in this experimental environment. 
Here we report the detection of neutrino interaction candidates in the pilot run data.

\section{The pilot run in LHC Run-2}

In 2018, we installed a 29 kg pilot detector in the TI18 tunnel, 480~m from the ATLAS IP, to measure neutrino interactions. With respect to the ATLAS IP, the TI18 tunnel is symmetric to TI12, where FASER$\nu$ will be located in LHC Run-3. Previously, we reported charged particle flux measurements made with other emulsion detectors installed in the TI12 and TI18 tunnels in 2018~\cite{Abreu:2020ddv}.  Here we focus on the pilot detector. 

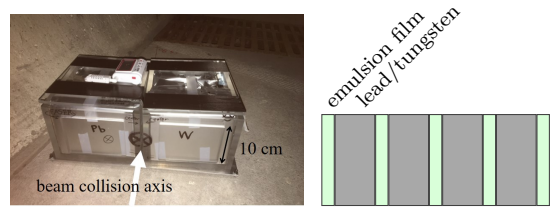
\begin{figure}[tb]
\centering
\includegraphics[width=0.45\linewidth]{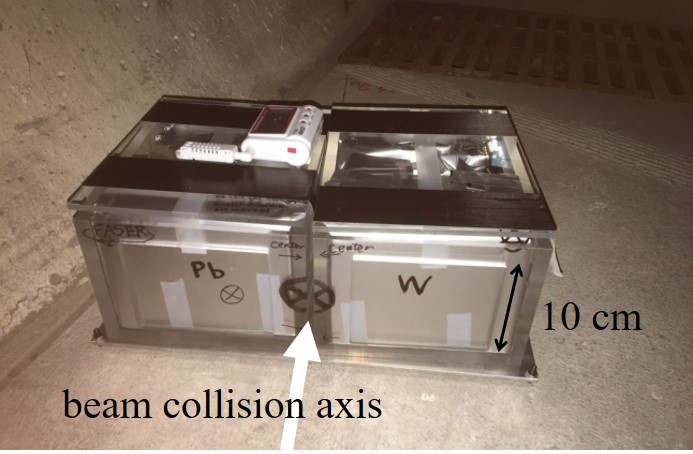}
\begin{tikzpicture}[scale=0.55]
\fill[green!15!white] (1.02,0) rectangle (1.3,2.2);
\draw [gray!50!black] (1.0,0) rectangle (1.3,2.2);
\foreach \i in {1,...,4}{
        \fill[gray!68!white] (1.3*\i+0.02,0) rectangle (1.3*\i+0.98,2.2);
        \draw [gray!50!black] (1.3*\i+0.02,0) rectangle (1.3*\i+0.98,2.2);
        \fill[green!15!white] (1.3*\i+1.0,0) rectangle (1.3*\i+1.3,2.2);
        \draw [gray!50!black] (1.3*\i+1.0,0) rectangle (1.3*\i+1.3,2.2);
    }
\node [rotate=45, anchor=west] at (1.,2.3) {\footnotesize{emulsion film}};
\node [rotate=45, anchor=west] at (1.8,2.3) {\footnotesize{lead/tungsten}};
\node [anchor=west] at (6.5,1.1) {{. . .}};
\end{tikzpicture}
\caption{Structure of the pilot emulsion detector. Metallic plates (1-mm-thick lead or 0.5-mm-thick tungsten) are interleaved with 0.3-mm-thick emulsion films. Only a schematic slice of the detector is depicted.
}
\label{fig:structure}
\end{figure}

The pilot detector is divided into a 14 kg module with 101 1-mm-thick lead plates and a 15 kg module with 120 0.5-mm-thick tungsten plates, each containing the corresponding number of emulsion films~\cite{Ariga2020}, as shown in Fig.~\ref{fig:structure}. These emulsion films and target plates were spare parts of the NA65/DsTau experiment~\cite{Aoki:2019jry}. Each module was vacuum-packed to preserve the alignment between films and placed in a 21-cm deep acrylic chamber. The transverse dimensions of the plates and films are 12.5 cm wide and 10 cm high.

The beam collision axis in TI18 was mapped out by the CERN survey team with mm precision. The two modules were placed side by side with the collision axis passing between them. The estimated uncertainty of the detector position is $\pm 1$ cm in both dimensions transverse to the collision axis. An integrated luminosity of 12.2 \si{fb^{-1}} with an uncertainty of 2\%, measured by the ATLAS experiment~\cite{ATLAS-CONF-2019-021,Avoni:2018iuv}, was collected during 4 weeks of data taking from September to October with $pp$ collisions at 13-TeV center-of-mass energy. The beam half-crossing-angle was about 150 $\mu$rad vertically upwards, which moves the collision axis at the FASER location upwards by $\sim$7 cm. The detector temperature was stable at 17.94 \si{^{\circ} C} with a standard deviation of 0.07 \si{^\circ C}~\cite{Ariga:2018pin}. Temperature stability is important to avoid displacement of the emulsion films and metallic plates and to ensure good alignment. The entire lead module and 15\% of the tungsten module were used in the following analysis; the remaining spare films in the tungsten module had data quality problems.

\section{Simulation}

Neutrinos produced in the forward direction at the LHC originate from the decay of hadrons, mainly pions, kaons, and $D$ mesons. Light hadron production is simulated using the \texttt{EPOS-LHC}~\cite{Pierog:2013ria}, \texttt{QGSJET-II-04}~\cite{Ostapchenko:2010vb}, \texttt{Sibyll~2.3c}~\cite{Ahn:2009wx, Riehn:2015oba}, and \texttt{DPMJET-III 2017.1}~\cite{Roesler:2000he, Fedynitch:2231593} simulation tools, as implemented in the \texttt{CRMC}~\cite{CRMC} package, while heavy hadron production is simulated using \texttt{Sibyll~2.3c}, \texttt{DPMJET-III 2017.1} and \texttt{Pythia~8.2}~\cite{Sjostrand:2006za, Sjostrand:2014zea} with the Monash tune~\cite{Skands:2014pea}. Long-lived hadrons are then propagated through the forward LHC beam pipe and magnetic fields using a dedicated simulation~\cite{nufluxes} implemented as a \texttt{Rivet} module~\cite{Bierlich:2019rhm}, using the geometry and beam optics for Run-2 as modeled by BDSIM~\cite{Nevay:2018zhp}. We use 13 TeV collision energy and a beam half-crossing-angle of 150~$\mu$rad vertically upwards. 
The hadrons are decayed at multiple locations along their trajectory according to decay branching fractions and kinematics provided by \texttt{Pythia~8.2}, and the spectra of neutrinos passing through the pilot detector are tabulated. We then use \texttt{Genie}~\cite{Andreopoulos:2009rq, Andreopoulos:2015wxa} with the configuration outlined in Ref.~\cite{Abreu:2019yak} to simulate neutrino interactions.

\begin{figure}[tb]
\centering
\includegraphics[width=0.80\linewidth]{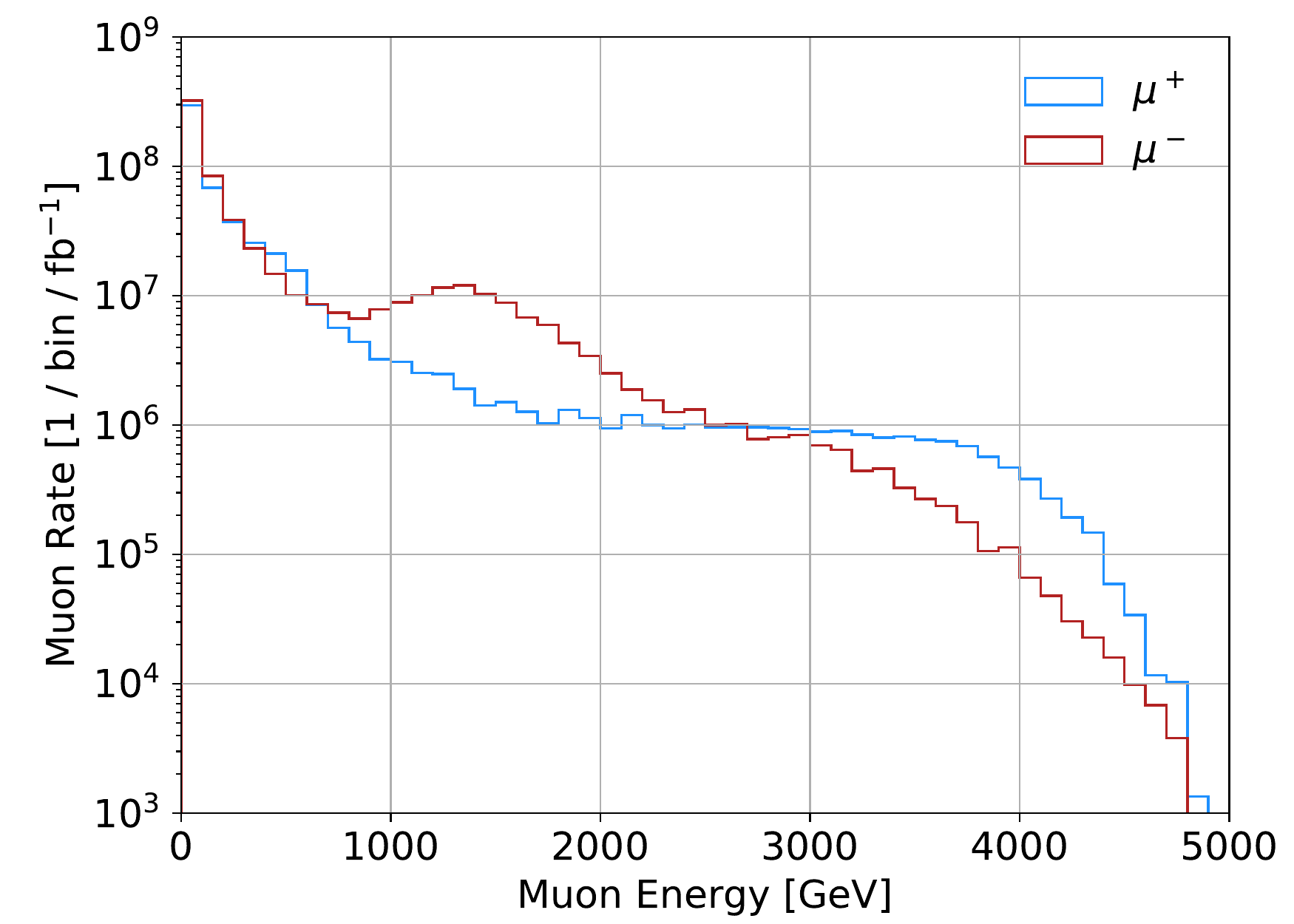}
\caption{The muon flux as a function of energy at 409 m from the IP, as estimated by FLUKA. Muons entering 1 m $\times$ 1 m around the collision axis are shown.}
\label{fig:muon_spectrum}
\end{figure}

The dominant source of background to neutrino interactions in the pilot run is inelastic interactions of neutral hadrons produced in muon photonuclear interactions upstream of the detector. The flux and spectrum of muons have been estimated by the CERN sources, targets, and interactions group, which performed FLUKA simulations~\cite{Ferrari:2005zk, Battistoni:2015epi}. 10$^8$ $pp$ collisions were simulated, and muons were propagated to the location when the beam collision axis leaves the concrete lining of the LHC tunnel (409 m from the IP) by the FLUKA simulation. The estimated muon flux as a function of energy at the 409 m position is shown in Fig.~\ref{fig:muon_spectrum}. The expected uncertainty on the FLUKA flux is of the order of 50\%. The muons were further propagated through 67 m of rock to reach close to the pilot detector by a \texttt{Geant4} simulation~\cite{AGOSTINELLI2003250}. The expected muon fluxes at the pilot detector position are $9.4\times10^3$ \si{\mu^-\!/cm^2/fb^{-1}} and $3.9\times10^3$ \si{\mu^+\!/cm^2/fb^{-1}} for $E_\mu > 100~\gev$, and $1.5\times10^4$ \si{\mu^-\!/cm^2/fb^{-1}} and $9.3\times10^3$ \si{\mu^+\!/cm^2/fb^{-1}} for $E_\mu>10~\gev$.

To simulate background hadron production and interactions, \texttt{Geant4} simulations of muons passing through the last 8 m of rock before reaching the pilot detector were performed. 
Neutral hadrons produced in the last 2 m in front of the pilot detector are the most relevant, because those produced further upstream are absorbed in the rock before reaching the detector.
The average rock density around CERN is measured to be about 2.5 g/cm$^3$~\cite{FERN2018249}. To reproduce this density, the rock was modeled as a mixture of 41\% CaCO$_3$ and 59\% SiO$_2$. 
10$^9$ negative muons and 10$^9$ positive muons were simulated. Kaons and neutrons are the relevant secondary neutral hadrons produced, with a small contribution from $\Lambda$ baryons. Table~\ref{tb:secodary} shows the production rates of neutral hadrons per incident muon for negative muons and positive muons. Muon-induced neutral hadrons have a steeply falling energy spectrum; a 10 GeV minimum energy threshold is applied to the simulation, since lower energy hadrons cannot satisfy the vertex reconstruction criteria used in our analysis. Neutral hadron interactions with the pilot detector were simulated by \texttt{Geant4} using the FTFP\_BERT and QGSP\_BERT physics lists, which correspond to different high-energy hadronic models~\cite{ALLISON2016186}.

\begin{table}[tb]
\caption{The production rates of neutral hadrons per incident muon with an energy threshold of 10 GeV. The difference between $\mu^-$ and $\mu^+$ is mainly due to the difference in the energy spectra.}
\vspace*{-.1in}
\begin{center}
\begin{ruledtabular}
\begin{tabular}{lcc}
& Negative Muons & Positive Muons \qquad {} \\ \hline
\rule{0pt}{3ex}\qquad $K_L$ & $3.3\times10^{-5}$  & $9.4\times10^{-6}$ \qquad {} \\
\qquad $K_S$                & $8.0\times10^{-6}$  & $2.3\times10^{-6}$ \qquad {} \\
\qquad $n$                  & $2.6\times10^{-5}$  & $7.7\times10^{-6}$ \qquad {} \\
\qquad $\bar{n}$            & $1.1\times10^{-5}$  & $3.2\times10^{-6}$ \qquad {} \\
\qquad $\Lambda$            & $3.5\times10^{-6}$  & $1.8\times10^{-6}$ \qquad {} \\
\qquad $\bar{\Lambda}$      & $2.8\times10^{-6}$  & $8.7\times10^{-7}$ \qquad {}
\end{tabular}
\end{ruledtabular}
\label{tb:secodary}
\end{center}
\end{table}

\section{Data analysis}

The data analysis is based on the readout of the full emulsion films by the Hyper Track Selector (HTS) system~\cite{Yoshimoto:2017ufm} with a readout speed of 0.45~m$^2$/hour/layer.
The HTS identifies track segments (``microtracks'') in the top and bottom emulsion layers of each film. A ``basetrack'' is formed by linking the two microtracks on a film. Each basetrack provides a 3D coordinate, 3D vector, and energy deposit ($dE/dx$) estimator.

Data processing is broken up into sub-volumes with a maximum size of 2 cm $\times$ 2 cm $\times$ 25 emulsion films. A preliminary alignment between each two consecutive films (position shifts and gap) is obtained using recorded tracks. To further improve the tracking resolution, an additional alignment calibration is applied by selecting tracks crossing many plates. Track reconstruction then links basetracks on different films by correlating their positions and angles. The track density in the data sample, some 10$^5$/cm$^2$ in a small angular space of 10 mrad, is relatively high compared to other emulsion experiments. A dedicated tracking algorithm for high density environments~\cite{Aoki:2019jry} is therefore employed on top of the software framework developed for the OPERA experiment \cite{Tyukov:2006ny}.

\begin{figure*}[tb]
    \begin{minipage}{0.67\textwidth}
    \begin{tabular}{cc}
    \begin{tikzpicture}
    \draw (0,0) node [anchor=north west, inner sep=0pt] {\includegraphics[width=0.5\linewidth]{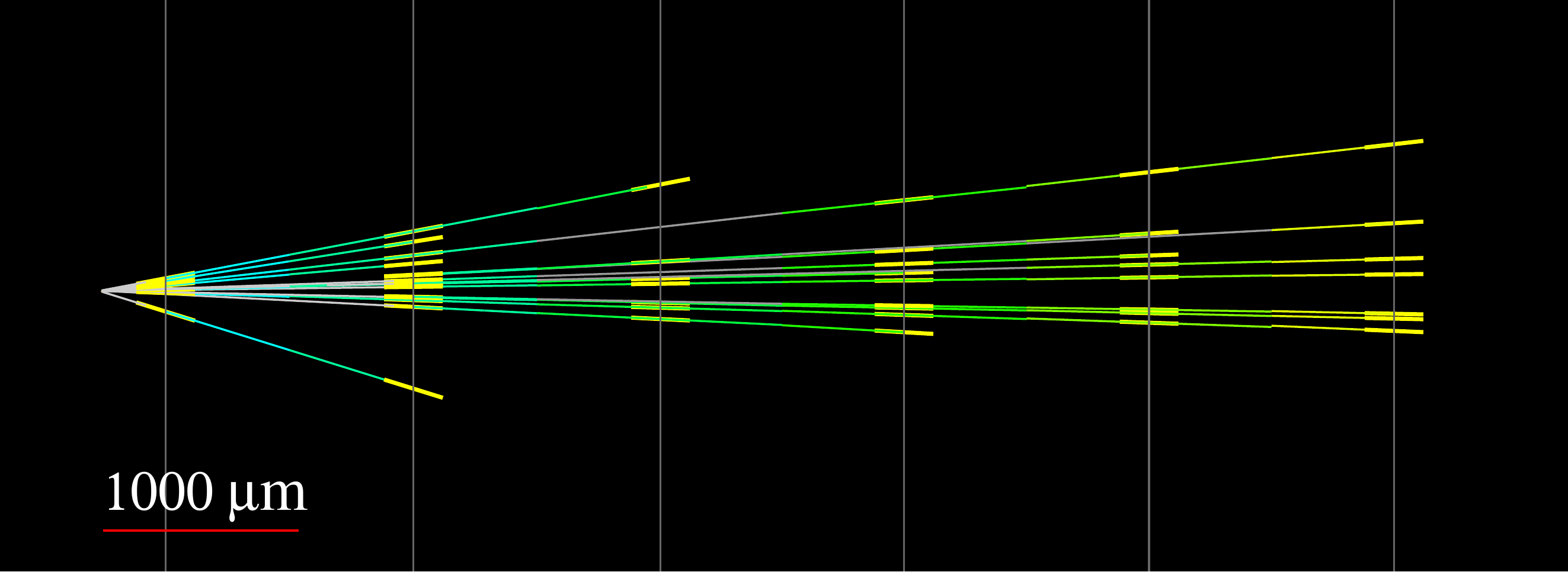}};
    \draw (0.05,-0.05) node [anchor=north west, inner sep=0pt] {\includegraphics[width=0.13\linewidth]{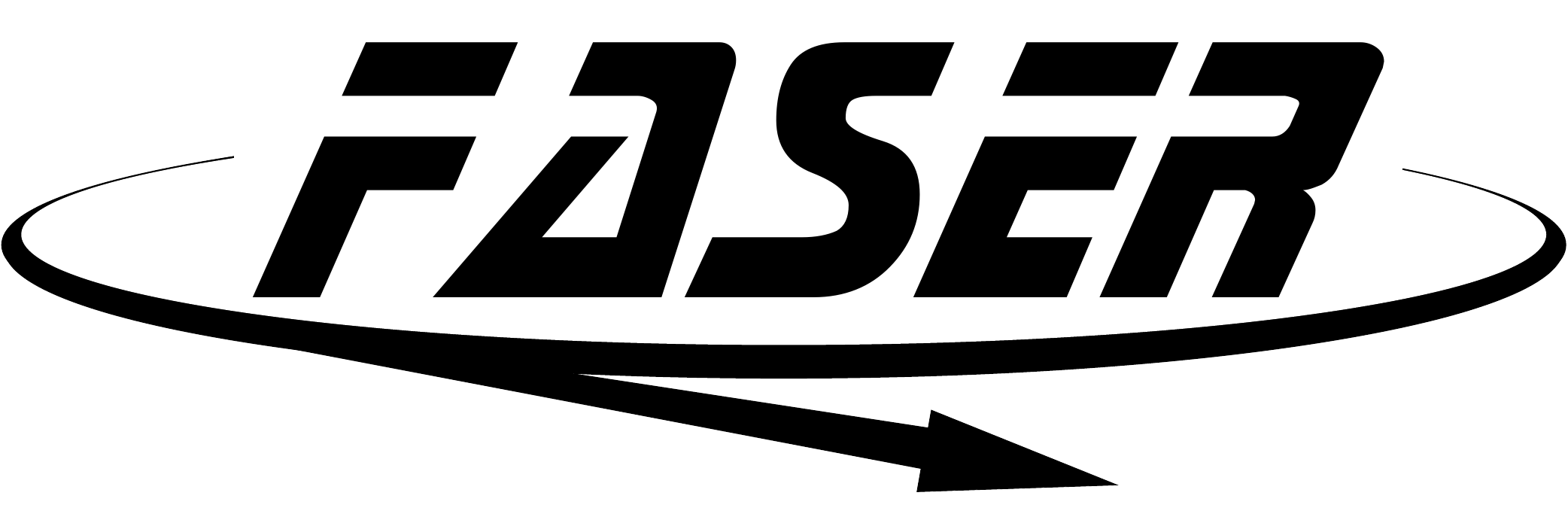}};
    \end{tikzpicture}
    &
    \includegraphics[width=0.4\linewidth]{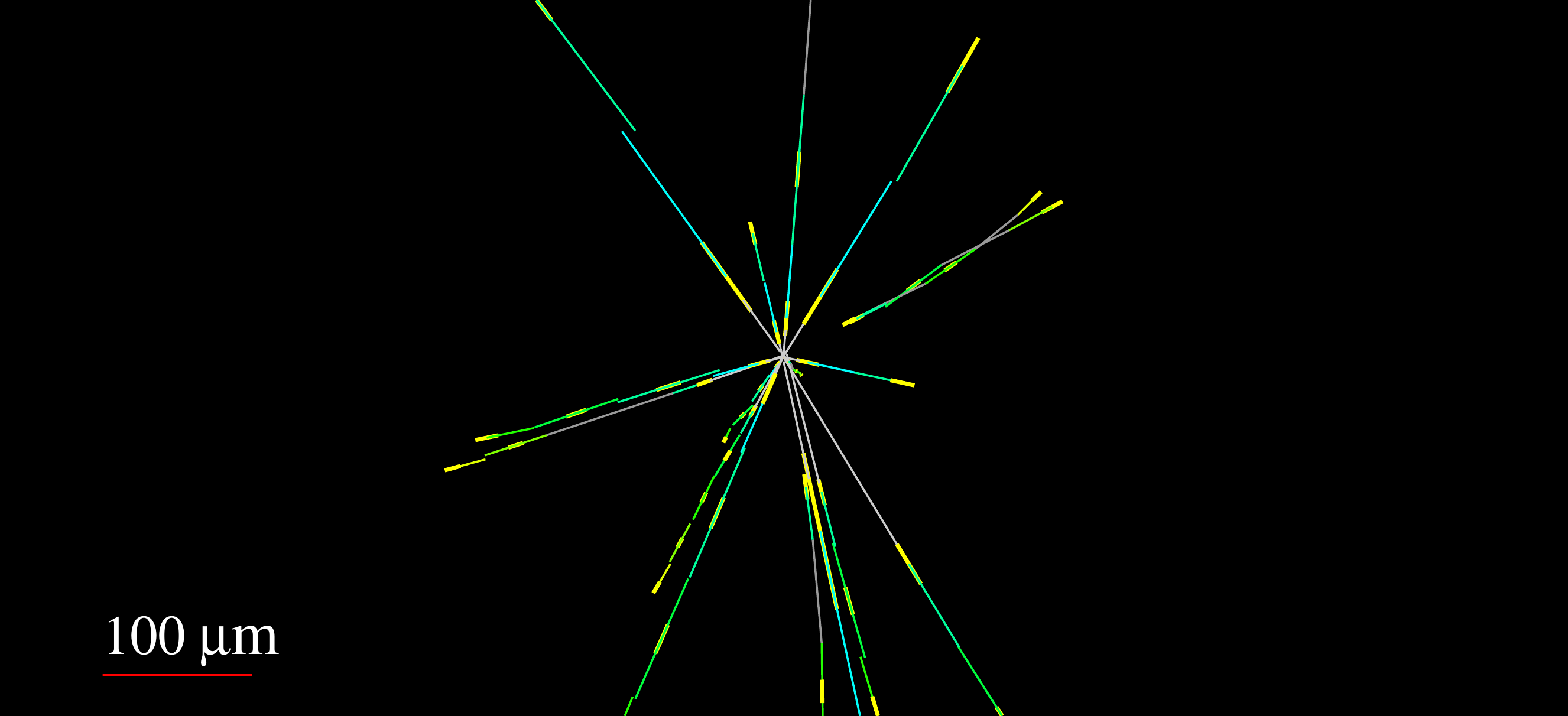}\\
    
    \begin{tikzpicture}
    \draw (0,0) node [anchor=north west, inner sep=0pt] {\includegraphics[width=0.5\linewidth]{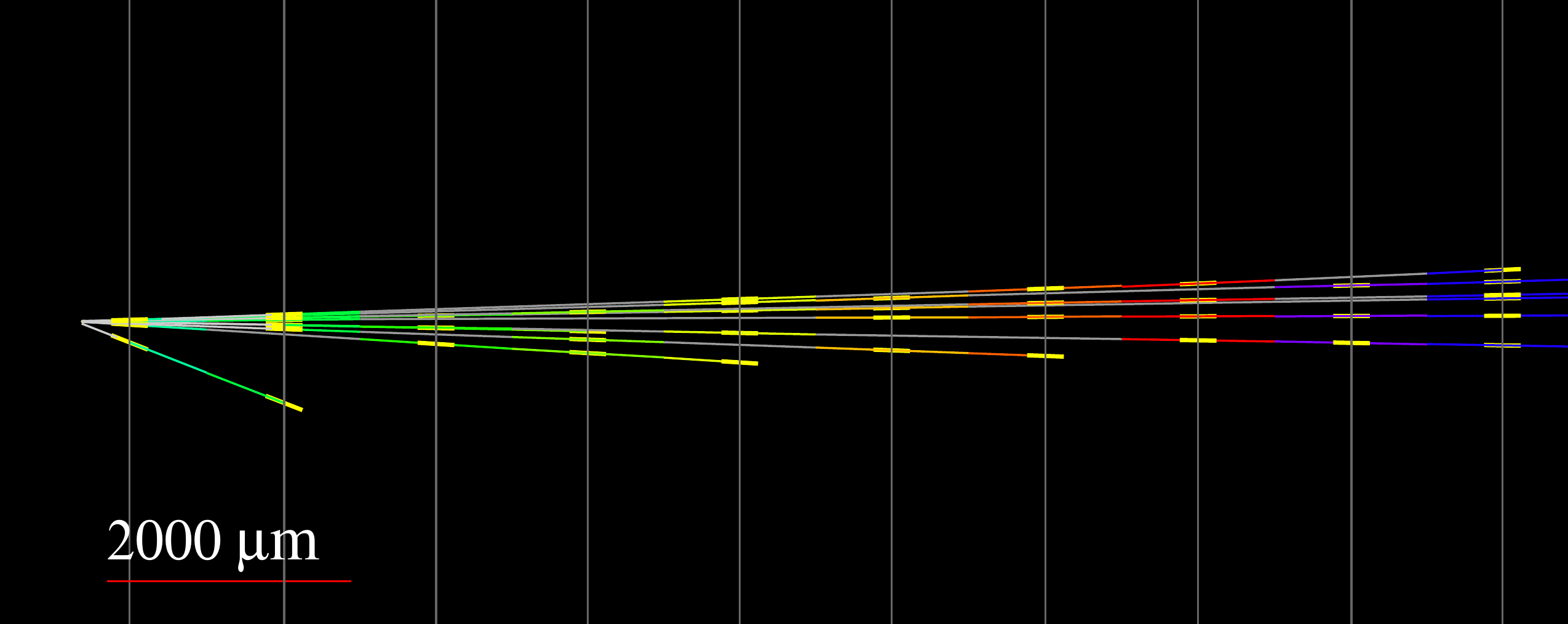}};
    \draw (0.05,-0.05) node [anchor=north west, inner sep=0pt] {\includegraphics[width=0.13\linewidth]{fig/FaserLogo.pdf}};
    \end{tikzpicture}
    &
\includegraphics[width=0.4\linewidth, trim  = 0 0 0 1.3cm, clip]{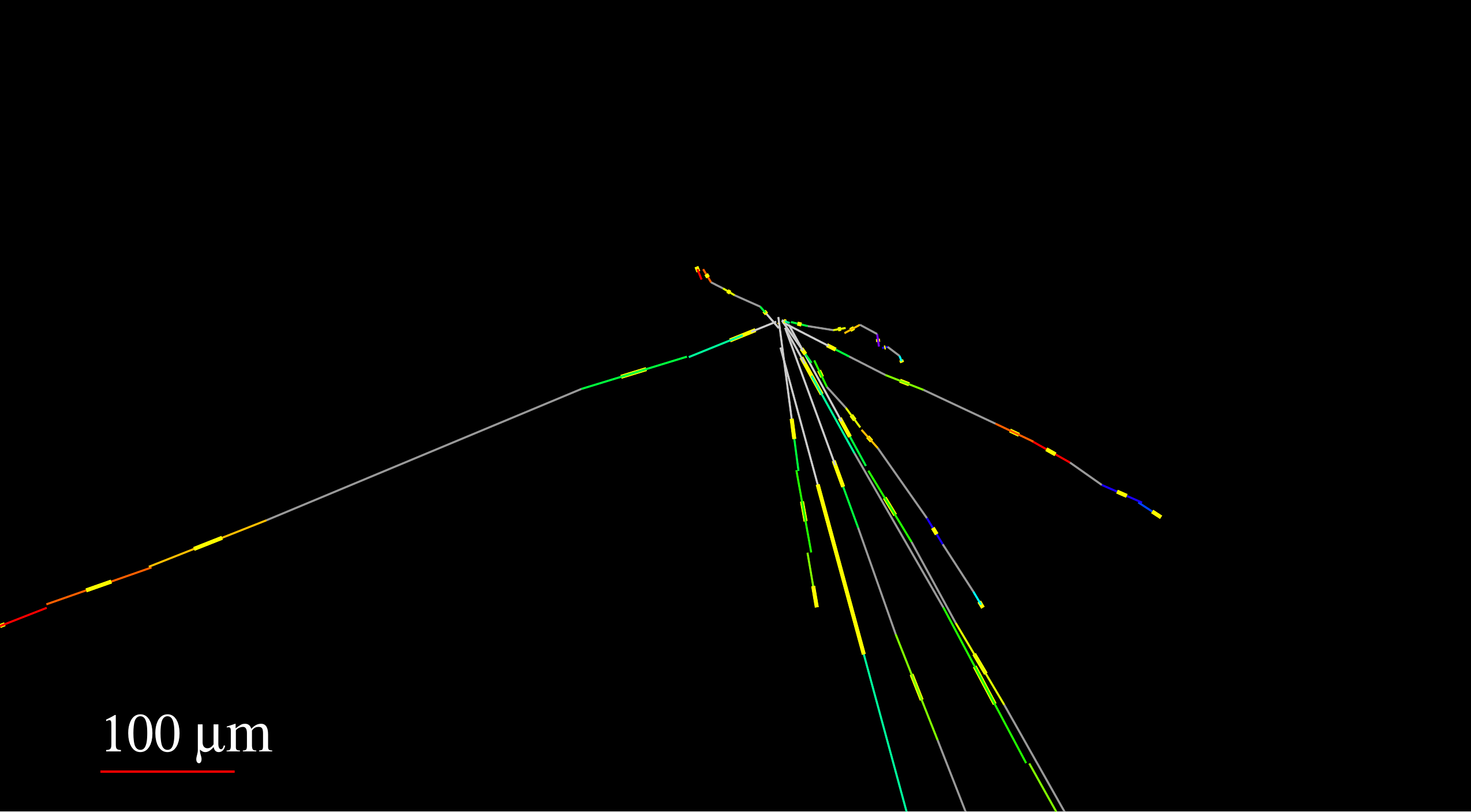}\\
    \end{tabular}
    \end{minipage}
    \begin{minipage}{0.29\textwidth}
\caption{Event displays of two of the neutral vertices in the $y$–$z$ projection longitudinal to the beam direction (left) and in the view transverse to the beam direction (right).
}
    \label{fig:vtx_Pb_area14_and_area11}
\end{minipage}
\end{figure*}

The majority of the tracks observed in the detector are expected to be background muons and related electromagnetic showers.
These background charged particles were analyzed using 10 emulsion films in the lead module. The position resolution in this data set is 0.5 $\mu$m and the angular resolution is 0.2 mrad. The observed angular distribution is peaked in the direction of the ATLAS IP. The angular spread of the peak is 2.3 mrad horizontally and 1.1 mrad vertically. The spatial distribution was uniform within the detector volume. Track detection efficiency was determined from the single film efficiency measured for tracks penetrating 10 plates. The estimated track detection efficiency for this flux measurement was $(88 \pm 5)$\%. After the efficiency correction, the charged particle flux within 10 mrad of the angular peak, which is dominated by energetic muons, is $(1.7 \pm 0.1) \times 10^4$ \si{tracks/cm^2/fb^{-1}} normalized by luminosity. This result is consistent with the values previously reported by other detectors~\cite{Ariga:2018pin,Abreu:2019yak} and close to the FLUKA prediction of $2.5\times10^4$ \si{tracks/cm^2/fb^{-1}} for $E_\mu>10$ GeV. 

For the neutrino analysis, using reconstructed tracks passing through at least 3 plates, vertex reconstruction was performed by searching for converging patterns of tracks with a minimum distance within 5 $\mu$m. Converging patterns with 5 or more tracks were then identified as vertices, rejecting the photon background. 
Collimation cuts were applied to these vertices to select high-energy interactions and suppress neutral hadron backgrounds: 
(1) the number of tracks with $\tan\theta \leq 0.1$ with respect to the beam direction is required to be 5 or more, and (2) the number of tracks with $\tan \theta > 0.1$ with respect to the beam direction is required to be 4 or less. Vertices are categorized as charged or neutral based on the presence or absence, respectively, of charged parent tracks. A looser track selection is used for the charged parent track search with a higher track detection efficiency of $99.8^{+0.1}_{-0.3}$\%. The background from charged vertices being reconstructed as neutral vertices is therefore negligible. The estimated selection efficiencies for neutrino signal and neutral hadron background vertices are shown in Table~\ref{tb:eff}. Signal classification is not performed in this analysis and interactions of all neutrino flavors are combined in the data.

\begin{table}[tb]
\caption{Efficiencies for selecting interaction vertices for the signal and background. The background efficiencies are estimated for interactions of neutral hadrons with energy $>$10~GeV. The statistical uncertainties are below 0.001 for all cases.
}
\begin{center}
\begin{ruledtabular}
\begin{tabular}{cc|ccc}
\multicolumn{2}{c|}{Signal} &\multicolumn{3}{c}{Background} \\ 
\hline
 & & & FTFP\_BERT& QGSP\_BERT \quad {} \\ 
 \hline
\quad $\nu_{e}$ & 0.490 \quad  & $K_L$ & 0.017 & 0.015  \quad {} \\
\quad $\bar{\nu_{e}}$ & 0.343 \quad & $K_S$ & 0.037 & 0.031 \quad {}  \\
\quad $\nu_{\mu}$ & 0.377 \quad  & $n$ & 0.011 & 0.012 \quad {}  \\
\quad $\bar{\nu_{\mu}}$  & 0.266 \quad  & $\bar{n}$ & 0.013 & 0.013 \quad {}  \\
\quad $\nu_{\tau}$ & 0.454 \quad  & $\Lambda$ & 0.020 & 0.021 \quad {}  \\
\quad $\bar{\nu_{\tau}}$ & 0.368 \quad & $\bar{\Lambda}$ & 0.018 & 0.018  \quad {} 
\end{tabular}
\end{ruledtabular}
\label{tb:eff}
\end{center}
\end{table}

The fiducial volume is defined by removing 7 films upstream, 5 films downstream, and 5 mm from the sides of the detector, corresponding to an 11 kg target mass. Within this volume, 18 neutral vertices passed the vertex selection criteria. 
Fig.~\ref{fig:vtx_Pb_area14_and_area11} shows two selected neutral vertices in lead, with 11 and 9 associated charged particles, respectively.

The expected number of neutrino signal vertices after all selections is $3.3^{+1.7}_{-0.9}$, dominated by muon neutrino interactions. The uncertainty reflects only the range of predictions obtained from different Monte Carlo simulations. The expected numbers of neutral hadron background vertices are 11.0 (FTFP\_BERT) and 10.1 (QGSP\_BERT). Since the difference of the two physics lists is not significant, FTFP\_BERT is used in the following analysis.

To validate the interaction features for the multivariate analysis described later, charged vertices (vertices with charged parent tracks attached), which simulation studies show also originate from muons, were analyzed. Although our muon flux measurement is close to the FLUKA prediction, no estimate of the uncertainty of the muon energy spectrum is available. The expected number of charged vertices satisfying the selection criteria is 115.4 (40.4 charged hadron interactions and 75.0 muon interactions), compared to the 78 charged vertices observed in the data.

\begin{figure*}[htb]
    \centering
    \includegraphics[width=0.9\linewidth]{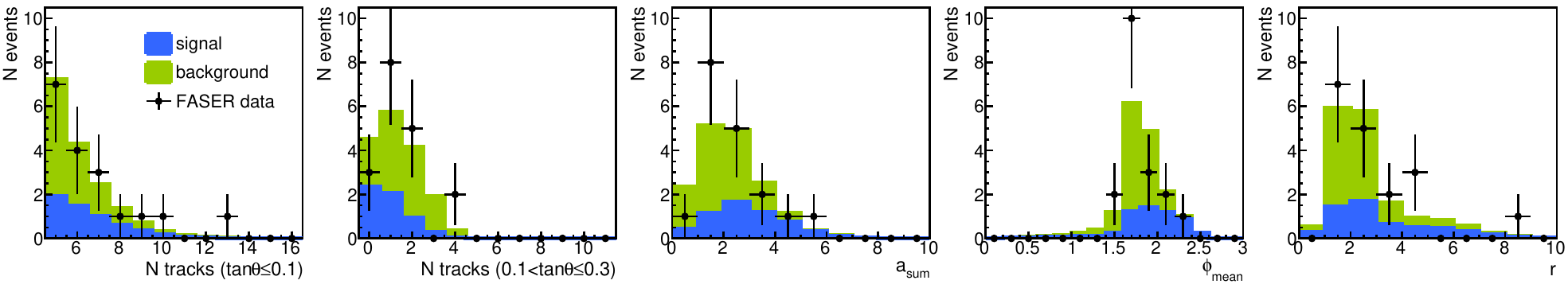}
    \caption{Monte Carlo simulation distributions of the BDT input variables for the neutrino signal and neutral hadron background. The observed neutral vertices in the data sample are shown in black. The Monte Carlo simulation distributions are normalized to 12.2 fb$^{-1}$.
    }
    \label{fig:dist_neutral}
\end{figure*}

\begin{figure*}[htb]
    \centering
    \includegraphics[width=0.9\linewidth]{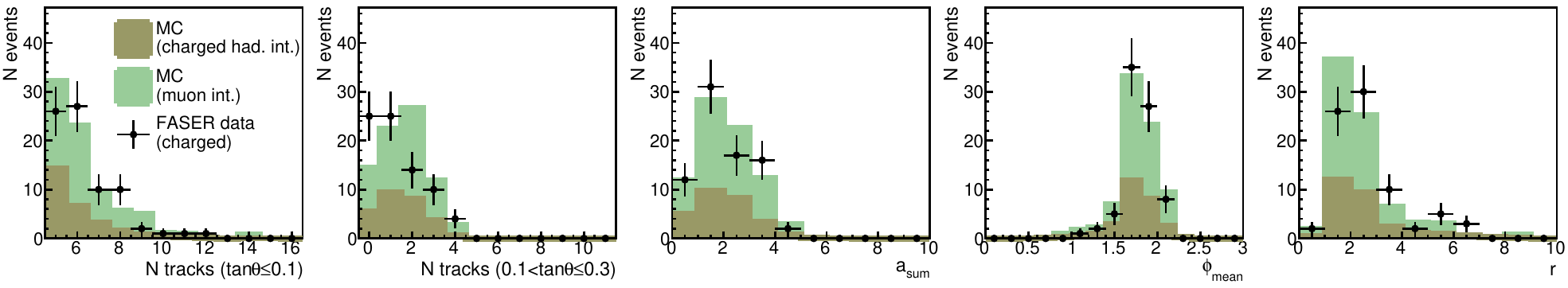}
    \caption{Monte Carlo simulation distributions of the BDT input variables for charged hadron interactions and muon interactions. The observed charged vertices in the data sample are shown in black. The Monte Carlo simulation distributions are normalized to the data to compare the shapes.
    }
    \label{fig:dist_charged}
\end{figure*}

\begin{figure}[htb]
    \centering
    \includegraphics[width=0.75\linewidth]{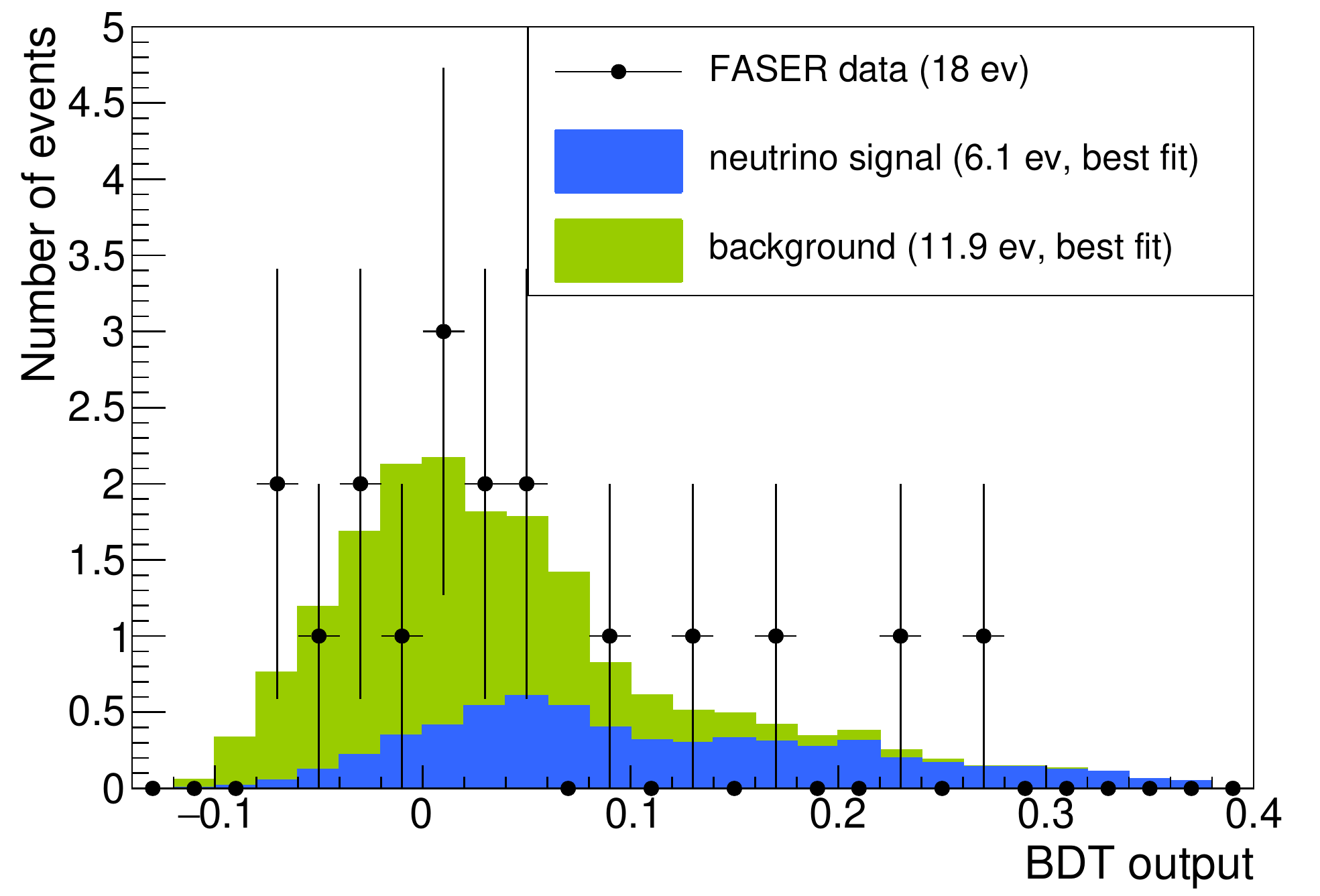}
    \caption{The BDT outputs of the observed neutral vertices, and the expected signal and background distributions (stacked) fitted to data. Higher BDT output values are associated with neutrino-like vertex features.
    }
    \label{fig:bdt_response}
\end{figure}

Since the pilot detector lacked the ability to identify muons, which could allow a clean separation of neutrino charged-current and neutral hadron vertices, we introduced the following multivariate approach as a much less powerful, but necessary, alternative. 
A multivariate discriminant, based on a boosted decision tree (BDT) algorithm, has been developed to distinguish neutrino signal from neutral hadron background in the neutral vertex sample. The BDT was implemented using the Toolkit for Multivariate Data Analysis~\cite{Hocker:2007ht} and trained with Monte Carlo events passing the vertex selection criteria. 
To define input variables for the BDT analysis, we selected high-energy interactions and checked the momentum balance of particles in the transverse plane. 
Using track angles with respect to the collision axis ($\theta$), the following variables were defined: (1) the number of tracks with $\tan\theta \leq 0.1$ with respect to the beam direction, (2) the number of tracks with $0.1 < \tan\theta \leq 0.3$ with respect to the beam direction, (3) the absolute value of the vector sum of transverse angles calculated considering all the tracks as unit vectors in the plane transverse to the beam direction ($a_{\text{sum}}$), (4) for each track in the event, we calculate the mean azimuthal angle between that track and all others, in the plane transverse to the beam direction, and then take the maximum value in the event ($\phi_{\text{mean}}$), (5) for each track in the event, we calculate the ratio of the number of tracks with azimuthal opening angle $\leq 90^\circ$ and $> 90^\circ$ in the plane transverse to the beam direction, and then take the maximum value in the event ($r$). 
The expected distributions of the input variables for the neutrino signal and for the neutral hadron background, compared with the data, are shown in Fig.~\ref{fig:dist_neutral}.

The charged vertices mentioned above can be used to validate the modeling of the BDT input variables in simulated data. Fig.~\ref{fig:dist_charged} shows that the BDT inputs for simulated charged hadron and muon interactions agree well with the charged vertex data.

The BDT estimator values for data and simulated neutral vertices are compared in Fig.~\ref{fig:bdt_response}. 
Here, the normalisation of the signal and background distributions is freely fitted to data, resulting in the best fit values of 6.1 and 11.9 events, respectively. The vertices shown in Fig.~\ref{fig:vtx_Pb_area14_and_area11} correspond to the first and second largest BDT values. An excess of events over the background expectation is observed at high BDT estimator, which is in agreement with the background plus signal hypothesis. A hypothesis test using the RooStats tool implemented in the CERN ROOT framework~\cite{fons_rademakers_2018_1292566} is carried out on the binned BDT estimator distribution. The background-only hypothesis is disfavored with a statistical significance of 2.7$\sigma$. The expected significance is estimated with pseudo experiments with the signal expectation of 3.3 events to be 1.7$\sigma$.

A systematic uncertainty related to the shape of the BDT distribution for neutrino events was estimated by varying the generator used for neutrino production and redoing the analysis. This resulted in a small ($<$0.2 events) change in the fitted neutrino yield.

Systematic uncertainties on the shape of the background BDT distribution were also evaluated by varying the shape of the muon energy distribution, by varying the modeling of the photonuclear interactions in the rock that produce the background neutral hadrons from the incoming muons, and varying the physics lists for the hadron interactions. These effects can change the energy and type of the neutral hadron interacting in the detector and therefore can influence the shape of the BDT distributions. The muon distribution was scaled up and down by a factor (1+$E$/3 TeV) distorting the spectrum as a function of energy, and the analysis repeated. Fitting the data with the updated background BDT shapes changed the fitted neutrino yield by 0.1 events. In addition, the analysis was repeated using FLUKA to model the production of neutral hadrons instead of \texttt{Geant4}, this leads to a change in the fitted neutrino yield of 0.1 events. Also, the analysis was repeated using the physics list QGSP\_BERT to model the hadron interactions instead of FTFP\_BERT, this leads to a change in the fitted neutrino yield of 0.1 events.

\section{Conclusions and outlook}

A search for neutrino interactions is presented based on a small emulsion detector installed at the LHC in 2018. We observe the first candidate vertices consistent with neutrino interactions at the LHC. A 2.7$\sigma$ excess of neutrino-like signal above muon-induced backgrounds is measured. 
These results demonstrate FASER$\nu$'s ability to detect neutrinos at the LHC and pave the way for future collider neutrino experiments.

We are currently preparing for data taking in LHC Run-3. With a deeper detector and lepton identification capability, FASER$\nu$ will perform better than the pilot run detector. In addition, the FASER spectrometer will measure the muon flux, reducing uncertainties on background estimates. 
In the 2022--2024 run, we expect to collect $\sim$10,000 flavor-tagged charged-current neutrino interactions.

\begin{acknowledgments}


We thank CERN for the excellent performance of the LHC and the technical and administrative staff members at all FASER institutions. We also gratefully acknowledge invaluable assistance from many groups at CERN, particularly the Physics Beyond Colliders study group; the ATLAS Collaboration for providing the luminosity value; the NA65/DsTau Collaboration for providing their spare emulsion films and tungsten plates for this measurement, and Masahiro Komatsu for useful discussions. This work was supported in part by Heising-Simons Foundation Grant Nos.~2018-1135, 2019-1179, and 2020-1840, Simons Foundation Grant No.~623683, and the Department of Energy Grant No.~DE-SC0016013. This work was supported by JSPS KAKENHI Grant Nos.~JP19H01909, JP20H01919, JP20K04004, JP20K23373, 
a research grant from the Mitsubishi Foundation, and the joint research program of the Institute of Materials and Systems for Sustainability.


\end{acknowledgments}


\bibliography{apssamp}

\providecommand{\href}[2]{#2}\begingroup\raggedright\begin{thebibliography}{10}

\bibitem{DeRujula:1984pg}
A.~De~Rujula and R.~Ruckl,
  \href{http://dx.doi.org/10.5170/CERN-1984-010-V-2.571}{``{Neutrino and muon
  physics in the collider mode of future accelerators},''} in {\em {SSC
  Workshop: Superconducting Super Collider Fixed Target Physics}},
  pp.~571--596.
\newblock 5, 1984.

\bibitem{Winter:1990ry}
K.~Winter, ``{Detection of the tau-neutrino at the LHC},'' in {\em {ECFA Large
  Hadron Collider (LHC) Workshop: Physics and Instrumentation}}, pp.~37--49.
\newblock 1990.

\bibitem{Vannucci:1993ud}
F.~Vannucci, ``{Neutrino physics at LHC / SSC},'' in {\em {4th International
  Symposium on Neutrino Telescopes}}, pp.~57--68.
\newblock 3, 1993.

\bibitem{DeRujula:1992sn}
A.~De~Rujula, E.~Fernandez, and J.~Gomez-Cadenas, ``{Neutrino fluxes at future
  hadron colliders},''
  \href{http://dx.doi.org/10.1016/0550-3213(93)90427-Q}{{\em Nucl. Phys. B}
  {\bf 405} (1993)  80--108}.

\bibitem{Park:2011gh}
H.~Park, ``{The estimation of neutrino fluxes produced by proton-proton
  collisions at $\sqrt{s}=14$ TeV of the LHC},''
  \href{http://dx.doi.org/10.1007/JHEP10(2011)092}{{\em JHEP} {\bf 10} (2011)
  092}, \href{http://arxiv.org/abs/1110.1971}{{\tt arXiv:1110.1971 [hep-ex]}}.

\bibitem{Feng:2017uoz}
J.~L. Feng, I.~Galon, F.~Kling, and S.~Trojanowski, ``{ForwArd Search
  ExpeRiment at the LHC},''
  \href{http://dx.doi.org/10.1103/PhysRevD.97.035001}{{\em Phys. Rev. D} {\bf
  97} (2018) no.~3, 035001}, \href{http://arxiv.org/abs/1708.09389}{{\tt
  arXiv:1708.09389 [hep-ph]}}.

\bibitem{Abreu:2019yak}
{\bf FASER} Collaboration, H.~Abreu {\em et al.}, ``{Detecting and Studying
  High-Energy Collider Neutrinos with FASER at the LHC},''
  \href{http://dx.doi.org/10.1140/epjc/s10052-020-7631-5}{{\em Eur. Phys. J. C}
  {\bf 80} (2020) no.~1, 61}, \href{http://arxiv.org/abs/1908.02310}{{\tt
  arXiv:1908.02310 [hep-ex]}}.

\bibitem{Zyla:2020zbs}
{\bf Particle Data Group} Collaboration, P.~A. Zyla {\em et al.}, ``{Review of
  Particle Physics},'' \href{http://dx.doi.org/10.1093/ptep/ptaa104}{{\em PTEP}
  {\bf 2020} (2020) no.~8, 083C01}.

\bibitem{Aartsen:2017kpd}
{\bf IceCube} Collaboration, M.~Aartsen {\em et al.}, ``{Measurement of the
  multi-TeV neutrino cross section with IceCube using Earth absorption},''
  \href{http://dx.doi.org/10.1038/nature24459}{{\em Nature} {\bf 551} (2017)
  596--600}, \href{http://arxiv.org/abs/1711.08119}{{\tt arXiv:1711.08119
  [hep-ex]}}.

\bibitem{Ahdida:2750060}
C.~Ahdida {\em et al.}, ``{SND@LHC - Scattering and Neutrino Detector at the
  LHC},'' tech. rep., CERN, Geneva, Jan, 2021.
\newblock \url{https://cds.cern.ch/record/2750060}.

\bibitem{Abreu:2020ddv}
{\bf FASER} Collaboration, H.~Abreu {\em et al.}, ``{Technical Proposal:
  FASERnu},'' \href{http://arxiv.org/abs/2001.03073}{{\tt arXiv:2001.03073
  [physics.ins-det]}}.

\bibitem{Ariga2020}
A.~Ariga, T.~Ariga, G.~D. Lellis, A.~Ereditato, and K.~Niwa, {\em Nuclear
  Emulsions},
  \href{http://dx.doi.org/10.1007/978-3-030-35318-6_9}{pp.~383--438}.
\newblock Springer International Publishing, Cham, 2020.
\newblock \url{https://doi.org/10.1007/978-3-030-35318-6_9}.

\bibitem{Aoki:2019jry}
{\bf DsTau} Collaboration, S.~Aoki {\em et al.}, ``{DsTau: Study of tau
  neutrino production with 400 GeV protons from the CERN-SPS},''
  \href{http://dx.doi.org/10.1007/JHEP01(2020)033}{{\em JHEP} {\bf 01} (2020)
  033}, \href{http://arxiv.org/abs/1906.03487}{{\tt arXiv:1906.03487
  [hep-ex]}}.

\bibitem{ATLAS-CONF-2019-021}
{\bf ATLAS} Collaboration, ``{Luminosity determination in $pp$ collisions at
  $\sqrt{s}=13$ TeV using the ATLAS detector at the LHC},'' Tech. Rep.
  ATLAS-CONF-2019-021, CERN, Geneva, Jun, 2019.
\newblock \url{https://cds.cern.ch/record/2677054}.

\bibitem{Avoni:2018iuv}
G.~Avoni {\em et al.}, ``{The new LUCID-2 detector for luminosity measurement
  and monitoring in ATLAS},''
  \href{http://dx.doi.org/10.1088/1748-0221/13/07/P07017}{{\em JINST} {\bf 13}
  (2018) no.~07, P07017}.

\bibitem{Ariga:2018pin}
{\bf FASER} Collaboration, A.~Ariga {\em et al.}, ``{Technical Proposal for
  FASER: ForwArd Search ExpeRiment at the LHC},''
  \href{http://arxiv.org/abs/1812.09139}{{\tt arXiv:1812.09139
  [physics.ins-det]}}.

\bibitem{Pierog:2013ria}
T.~Pierog, I.~Karpenko, J.~M. Katzy, E.~Yatsenko, and K.~Werner, ``{EPOS LHC:
  Test of collective hadronization with data measured at the CERN Large Hadron
  Collider},'' \href{http://dx.doi.org/10.1103/PhysRevC.92.034906}{{\em Phys.
  Rev.} {\bf C92} (2015)  034906},
\href{http://arxiv.org/abs/1306.0121}{{\tt arXiv:1306.0121 [hep-ph]}}.

\bibitem{Ostapchenko:2010vb}
S.~Ostapchenko, ``{Monte Carlo treatment of hadronic interactions in enhanced
  Pomeron scheme: I. QGSJET-II model},''
  \href{http://dx.doi.org/10.1103/PhysRevD.83.014018}{{\em Phys. Rev.} {\bf
  D83} (2011)  014018},
\href{http://arxiv.org/abs/1010.1869}{{\tt arXiv:1010.1869 [hep-ph]}}.

\bibitem{Ahn:2009wx}
E.-J. Ahn, R.~Engel, T.~K. Gaisser, P.~Lipari, and T.~Stanev, ``{Cosmic ray
  interaction event generator SIBYLL 2.1},''
  \href{http://dx.doi.org/10.1103/PhysRevD.80.094003}{{\em Phys. Rev.} {\bf
  D80} (2009)  094003},
\href{http://arxiv.org/abs/0906.4113}{{\tt arXiv:0906.4113 [hep-ph]}}.

\bibitem{Riehn:2015oba}
F.~Riehn, R.~Engel, A.~Fedynitch, T.~K. Gaisser, and T.~Stanev, ``{A new
  version of the event generator Sibyll},'' {\em PoS} {\bf ICRC2015} (2016)
  558,
\href{http://arxiv.org/abs/1510.00568}{{\tt arXiv:1510.00568 [hep-ph]}}.

\bibitem{Roesler:2000he}
S.~Roesler, R.~Engel, and J.~Ranft,
  \href{http://dx.doi.org/10.1007/978-3-642-18211-2_166}{``{The Monte Carlo
  event generator DPMJET-III},''} in {\em {International Conference on Advanced
  Monte Carlo for Radiation Physics, Particle Transport Simulation and
  Applications (MC 2000)}}.
\newblock 12, 2000.
\newblock \href{http://arxiv.org/abs/hep-ph/0012252}{{\tt
  arXiv:hep-ph/0012252}}.

\bibitem{Fedynitch:2231593}
A.~Fedynitch, ``{Cascade equations and hadronic interactions at very high
  energies},'' Nov, 2015.
\newblock \url{https://cds.cern.ch/record/2231593}. Presented 27 Nov 2015.

\bibitem{CRMC}
C.~Baus, T.~Pierog, and R.~Ulrich, ``{Cosmic Ray Monte Carlo (CRMC)},''.
  \url{https://web.ikp.kit.edu/rulrich/crmc.html}.

\bibitem{Sjostrand:2006za}
T.~Sjostrand, S.~Mrenna, and P.~Z. Skands, ``{PYTHIA 6.4 Physics and Manual},''
  \href{http://dx.doi.org/10.1088/1126-6708/2006/05/026}{{\em JHEP} {\bf 05}
  (2006)  026},
\href{http://arxiv.org/abs/hep-ph/0603175}{{\tt arXiv:hep-ph/0603175
  [hep-ph]}}.

\bibitem{Sjostrand:2014zea}
T.~Sjöstrand, S.~Ask, J.~R. Christiansen, R.~Corke, N.~Desai, P.~Ilten,
  S.~Mrenna, S.~Prestel, C.~O. Rasmussen, and P.~Z. Skands, ``{An Introduction
  to PYTHIA 8.2},'' \href{http://dx.doi.org/10.1016/j.cpc.2015.01.024}{{\em
  Comput. Phys. Commun.} {\bf 191} (2015)  159--177},
  \href{http://arxiv.org/abs/1410.3012}{{\tt arXiv:1410.3012 [hep-ph]}}.

\bibitem{Skands:2014pea}
P.~Skands, S.~Carrazza, and J.~Rojo, ``{Tuning PYTHIA 8.1: the Monash 2013
  Tune},'' \href{http://dx.doi.org/10.1140/epjc/s10052-014-3024-y}{{\em Eur.
  Phys. J.} {\bf C74} (2014) no.~8, 3024},
\href{http://arxiv.org/abs/1404.5630}{{\tt arXiv:1404.5630 [hep-ph]}}.

\bibitem{nufluxes}
F.~Kling, ``{Forward Neutrino Fluxes and Simulation at the LHC}.'' in
  preparation.

\bibitem{Bierlich:2019rhm}
C.~Bierlich {\em et al.}, ``{Robust Independent Validation of Experiment and
  Theory: Rivet version 3},''
  \href{http://dx.doi.org/10.21468/SciPostPhys.8.2.026}{{\em SciPost Phys.}
  {\bf 8} (2020)  026}, \href{http://arxiv.org/abs/1912.05451}{{\tt
  arXiv:1912.05451 [hep-ph]}}.

\bibitem{Nevay:2018zhp}
L.~J. Nevay {\em et al.}, ``{BDSIM: An accelerator tracking code with
  particle\textendash{}matter interactions},''
  \href{http://dx.doi.org/10.1016/j.cpc.2020.107200}{{\em Comput. Phys.
  Commun.} {\bf 252} (2020)  107200},
  \href{http://arxiv.org/abs/1808.10745}{{\tt arXiv:1808.10745
  [physics.comp-ph]}}.

\bibitem{Andreopoulos:2009rq}
C.~Andreopoulos {\em et al.}, ``{The GENIE Neutrino Monte Carlo Generator},''
  \href{http://dx.doi.org/10.1016/j.nima.2009.12.009}{{\em Nucl. Instrum.
  Meth.} {\bf A614} (2010)  87--104},
\href{http://arxiv.org/abs/0905.2517}{{\tt arXiv:0905.2517 [hep-ph]}}.

\bibitem{Andreopoulos:2015wxa}
C.~Andreopoulos, C.~Barry, S.~Dytman, H.~Gallagher, T.~Golan, R.~Hatcher,
  G.~Perdue, and J.~Yarba, ``{The GENIE Neutrino Monte Carlo Generator: Physics
  and User Manual},''
\href{http://arxiv.org/abs/1510.05494}{{\tt arXiv:1510.05494 [hep-ph]}}.

\bibitem{Ferrari:2005zk}
A.~Ferrari, P.~R. Sala, A.~Fasso, and J.~Ranft, ``{FLUKA: A multi-particle
  transport code (Program version 2005)},''.

\bibitem{Battistoni:2015epi}
G.~Battistoni {\em et al.}, ``{Overview of the FLUKA code},''
  \href{http://dx.doi.org/10.1016/j.anucene.2014.11.007}{{\em Annals Nucl.
  Energy} {\bf 82} (2015)  10--18}.

\bibitem{AGOSTINELLI2003250}
S.~Agostinelli {\em et al.}, ``Geant4 - simulation toolkit,''
  \href{http://dx.doi.org/https://doi.org/10.1016/S0168-9002(03)01368-8}{{\em
  Nuclear Instruments and Methods in Physics Research Section A: Accelerators,
  Spectrometers, Detectors and Associated Equipment} {\bf 506} (2003) no.~3,
  250 -- 303}.
  \url{http://www.sciencedirect.com/science/article/pii/S0168900203013688}.

\bibitem{FERN2018249}
E.~J. Fern, V.~{Di Murro}, K.~Soga, Z.~Li, L.~Scibile, and J.~A. Osborne,
  ``Geotechnical characterisation of a weak sedimentary rock mass at cern,
  geneva,''
  \href{http://dx.doi.org/https://doi.org/10.1016/j.tust.2018.04.003}{{\em
  Tunnelling and Underground Space Technology} {\bf 77} (2018)  249--260}.
  \url{http://www.sciencedirect.com/science/article/pii/S0886779817310349}.

\bibitem{ALLISON2016186}
J.~Allison {\em et al.}, ``{Recent developments in Geant4},''
  \href{http://dx.doi.org/https://doi.org/10.1016/j.nima.2016.06.125}{{\em
  Nuclear Instruments and Methods in Physics Research Section A: Accelerators,
  Spectrometers, Detectors and Associated Equipment} {\bf 835} (2016)  186 --
  225}.
  \url{http://www.sciencedirect.com/science/article/pii/S0168900216306957}.

\bibitem{Yoshimoto:2017ufm}
M.~Yoshimoto, T.~Nakano, R.~Komatani, and H.~Kawahara, ``{Hyper-track selector
  nuclear emulsion readout system aimed at scanning an area of one thousand
  square meters},'' \href{http://dx.doi.org/10.1093/ptep/ptx131}{{\em PTEP}
  {\bf 2017} (2017) no.~10, 103H01},
  \href{http://arxiv.org/abs/1704.06814}{{\tt arXiv:1704.06814
  [physics.ins-det]}}.

\bibitem{Tyukov:2006ny}
V.~Tyukov, I.~Kreslo, Y.~Petukhov, and G.~Sirri, ``{The FEDRA Framework for
  emulsion data reconstruction and analysis in the OPERA experiment},''
  \href{http://dx.doi.org/10.1016/j.nima.2005.11.214}{{\em Nucl. Instrum. Meth.
  A} {\bf 559} (2006)  103--105}.

\bibitem{Hocker:2007ht}
A.~Hoecker {\em et al.}, ``{TMVA - Toolkit for Multivariate Data Analysis},''
  \href{http://arxiv.org/abs/physics/0703039}{{\tt arXiv:physics/0703039}}.

\bibitem{fons_rademakers_2018_1292566}
F.~Rademakers {\em et al.}, ``root-project/root: v6.16/02,'' June, 2018.
\newblock \url{https://doi.org/10.5281/zenodo.1292566}.

\end{thebibliography}\endgroup


\providecommand{\href}[2]{#2}\begingroup\raggedright\endgroup

\end{document}